# QUANTUM ATOM OPTICS WITH BOSONS AND FERMIONS

Alain Aspect, Denis Boiron, Chris Westbrook
*Laboratoire Charles Fabry de l'Institut Optique, CNRS, Univ Paris-sud*
*Campus Polytechnique RD 128, 91127 Palaiseau Cedex*

Atom optics, a field which takes much inspiration from traditional optics, has advanced to the point that some of the fundamental experiments of quantum optics, involving photon correlations, have found atomic analogs. We discuss some recent experiments on atom bunching and anti-bunching as well as some prospects for extending them to the field of many body physics.

Most of the time, experimental physicists fight to eliminate noise. But the history of physics is full of examples in which the study of noise, or more precisely fluctuation phenomena, has led to significant discoveries. A famous example is Einstein's analysis of Brownian motion in 1905 which played a crucial role in the development of the atomic theory of matter. And several times in his later career, Einstein returned to the study of fluctuations. In the famous 1925 article in which he described what we now know as Bose-Einstein condensation[1], Einstein considered the fluctuation in the number of particles $N$ in a small volume within a larger volume of an ideal gas at a given temperature. Using thermodynamic arguments, he found a formula for the variance of $N$:

$$\delta N^2 = \langle N \rangle + \langle N \rangle^2 / g \qquad (1)$$

Recall that the variance is the mean squared deviation from the mean of $N$: $\delta N^2 = \langle (N - \langle N \rangle)^2 \rangle$. In Eq. (1) the quantity $g$ is the number of phase space cells occupied by the gas, that is the phase space volume divided by Planck's constant cubed: $g = (\Delta x \Delta p/h)^3$.

In reading the paper, one can sense his fascination with this formula: he identifies the linear term with the fluctuations of independent particles, what we often call shot noise today. Einstein recognizes the quadratic term as being due to an interference effect. Indeed he had already made a similar analysis in the case of radiation[2], in which case the quadratically increasing variance can be interpereted as "speckle" (see Box 1). Here however, the formula applies to matter and he remarks on the "mutual influence between the particles of an altogether puzzling nature". With the formulation of the quantum theory in the ensuing few years, it became clear that he had put his finger once again on the wave-particle duality.

This formula became standard fare in textbooks on quantum statistical mechanics[3], but like many calculations in textbooks, it applied to a *Gedankenexperiment*. The quadratic term is usually extremely small compared to the linear term, which itself is often difficult to observe. For example at atmospheric temperature and pressure, 1 mm$^3$ of air contains about 2 10$^{16}$ molecules. According to (1), the shot noise is of order 10$^{-8}$ and the interference term is another 10$^6$ times smaller. In other words the spatial and temporal coherence of a typical gas of atoms or photons is extremely small. To our knowledge the first observation of these interference fluctuations was made using light, in the famous experiments of Robert Hanbury Brown and Richard Twiss (HBT), which we shall describe below (see Fig. 1). The HBT experiment stimulated deep questions about the quantum description of light and gave rise to the birth of modern quantum optics. This line of research was recently recognized in the attribution of part of the Nobel prize for physics to Roy Glauber for his "contributions to the quantum theory of optical coherence".

Since the appearance of the first atom interferometers in the 1990's the field of atom optics has made tremendous progress, often inspired by traditional optics. Recently it has become possible to realize fluctuation experiments analogous to the HBT experiment. In the following we will describe some of these experiments. The first ones, carried out with bosons, give results highly analogous to those with photons. The results are of course unsurprising since photons are also bosons. Still, they nicely demonstrate this puzzling "mutual influence between particles", an influence which we have now learned to interpret as two particle quantum interference. In atom optics, one can also use fermions. The quadratic term in Eq. 1 appears with a minus sign, a manifestation of the exclusion principle, and it gives rise to an effect which has no optical or classical wave analog.

We will begin with a discussion of the Hanbury Brown – Twiss experiments using light. This discussion will allow us to introduce practically all the necessary concepts to understand the second part in which we discuss the case of atoms.

**The Hanbury Brown – Twiss Experiment**

First, a bit of the history of photon correlations. After his experience in the development of radar in the second world war, Hanbury Brown turned to radio astronomy. He proposed a new method to measure the angular diameter of a star by studying correlations in the fluctations in two radio telescopes as a function of their separation. The method depends on a generalization of Eq. 1 to fluctuations at two separate points in space (see Box.1). If two telescopes are close together, their fluctuations are correlated and the quadratic term in Eq. 1 is present. A measurement of the distance over which this correlation persists is related to the angular diameter of the star (see Box 1).

After demonstrating the method in the radio frequency domain, Hanbury Brown proposed an extension to the visible. The proposal was greeted with great skepticism. Perhaps the with recent introduction of photomultiplier tubes, based on the photoelectric effect, physicists had become accustomed to thinking of light in terms of photons. But the photon description of Hanbury Brown's method is to say the least, surprising. A correlation between the detection probabilities of different photons seemed to imply that the photons "know" that they should arrive together, even if they were emitted by two well separated points on a distant star. Box 2 gives a quantum explication of how this is possible. But in 1955 Hanbury Brown could not convince his colleagues nor funding agents of the validity of his idea.

To convince the skeptics Hanbury Brown began a collaboration with Twiss to perform a laboratory demonstration of the method[4] (see Fig. 1). Their system of two detectors observing the same source through a beam splitter is now a standard quantum optics tool. The results confirmed Hanbury Brown's idea and the two scientists made their first astronomical observation on the star Sirius, whose high brightness facilitated the observation. In subsequent work done in Australia, they measured the angular diameter of several other stars.

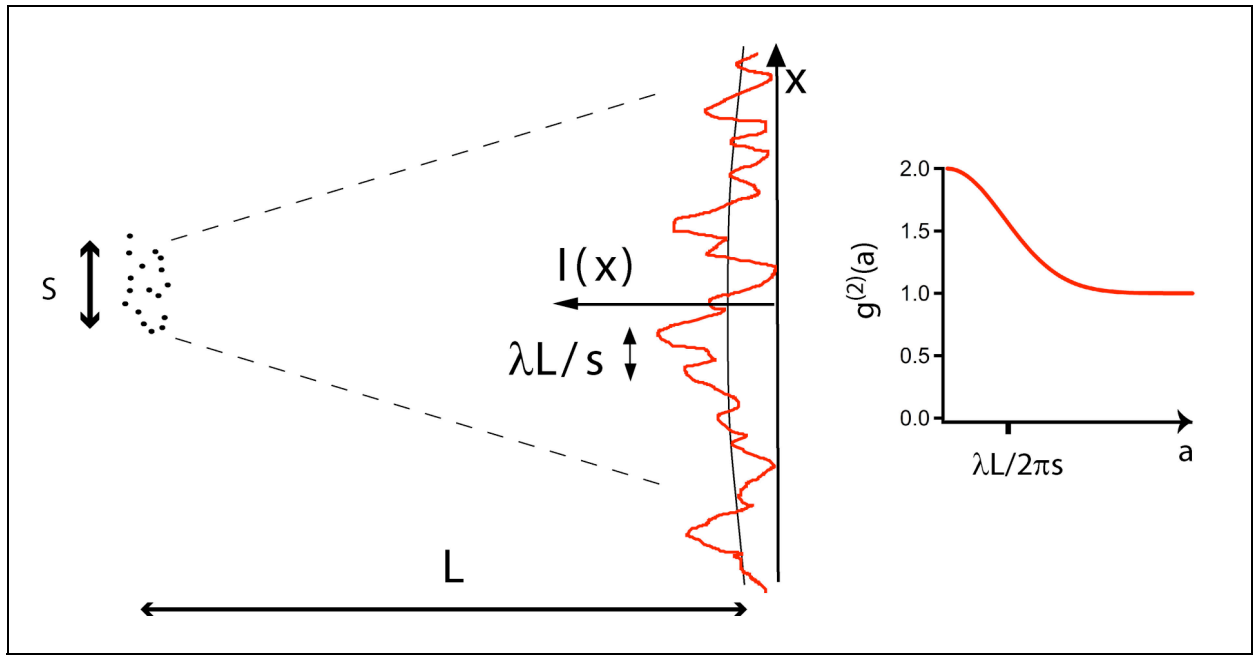

Box 1. Intensity correlations can be interpreted in terms of speckle. A source region of diameter $s$ contains a large number of independent (randomly phased) point sources. On a screen at a distance $L$, there appears a random interference pattern known as speckle. For two points $x$ and $x+a$ with a sufficient spatial separation, the intensities at a given time $t$ are not correlated and $\langle I(t,x)\,I(t,x+a)\rangle = \langle I\rangle^2$. For zero separation the intensities are correlated and therefore $\langle I^2\rangle \geq \langle I\rangle^2$. For thermal radiation, and neglecting shot noise, Einstein's formula reads: $\langle I^2\rangle = 2\langle I\rangle^2$, if we identify $I$ with the number of photons. The averages $\langle.\rangle$ can be thought of either as ensemble averages over a large number of equivalent realizations, or as time averages if one supposes that the relative phases of the source points are time dependent (for example because of atomic motion or collisions in a discharge lamp). One defines a correlation function $g^{(2)}(a) = \langle I(t,x)\,I(t,x+a)\rangle / \langle I(t,x)\rangle\,\langle I(t,x+a)\rangle$. The width of this function corresponds to the typical width of the speckles and represents the distance over which the relative phases associated with the optical paths between the source points and the observation points vary by less than 1 radian. The width is $\lambda L / 2\pi s$, and thus a measurement of this width leads to the angular size of the source $s / L$.

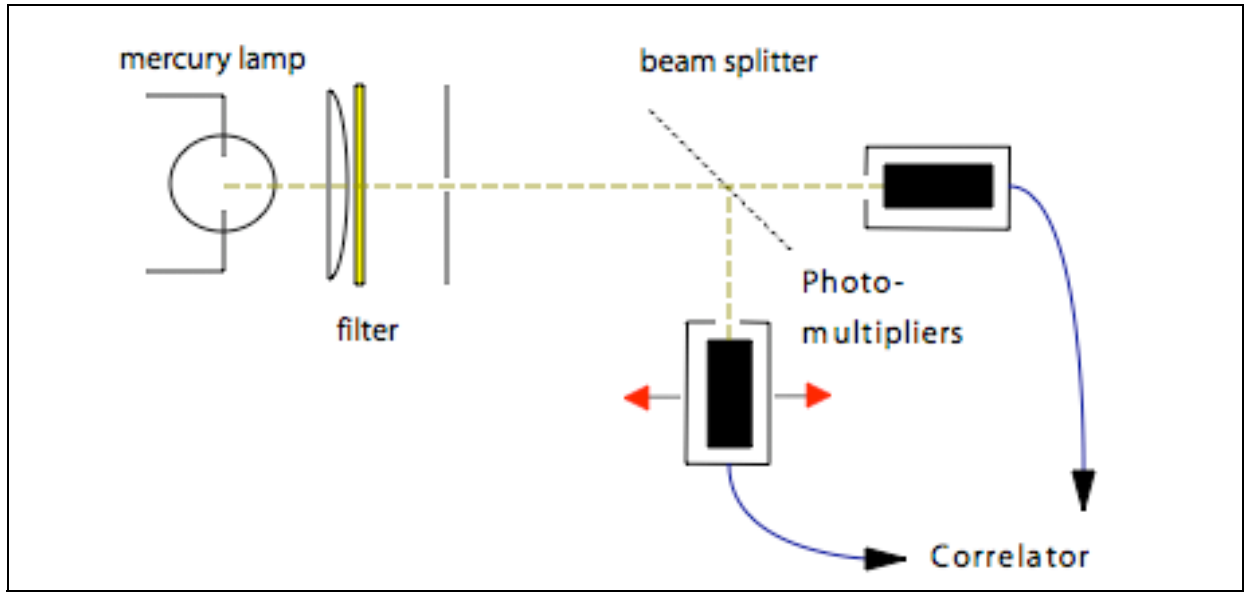

Figure 1. The Hanbury Brown – Twiss experiment. The source is the small slit illuminated by the mercury lamp. A filter selects a single emission line in order to lower the time scale of the fluctuations which are much faster than those of the detectors. The correlator multiplies the photo currents and averages them over time. If the image of one detector in the beam splitter is superposed with the other one, the photocurrents are correlated. A displacement of one of the detectors (indicated by the red arrows), corresponds to changing the distance *a* in Box 1. The correlation disappears when the detectors are separated by more than a correlation length. In terms of photons it means that they tend to be detected together or "bunched", a surprising effect for 2 photons emitted from two points on the surface of a star.

Despite the experiments, controversies continued. The invention of the laser in 1960 posed new questions. Debates raged about the highly coherent laser light: would it, exhibit HBT correlations? Glauber gave a clear answer in 1963 using quantized fields and a careful examination of what constitutes photon detection[5]. The result was that unlike "normal light", the photons emitted by a laser are in general *not* correlated (see Box 2). By 1965 experiments had confirmed Glauber's analysis. The quantum theory of optical coherence led experimenters to investigate situations in which photon correlations cannot be understood in terms of classical waves, unlike the HBT experiment. The development of methods to create and use new states of light (single photons, squeezed states and entangled states), not described by classical electromagnetism has been an essential occupation of the field of quantum optics since then.

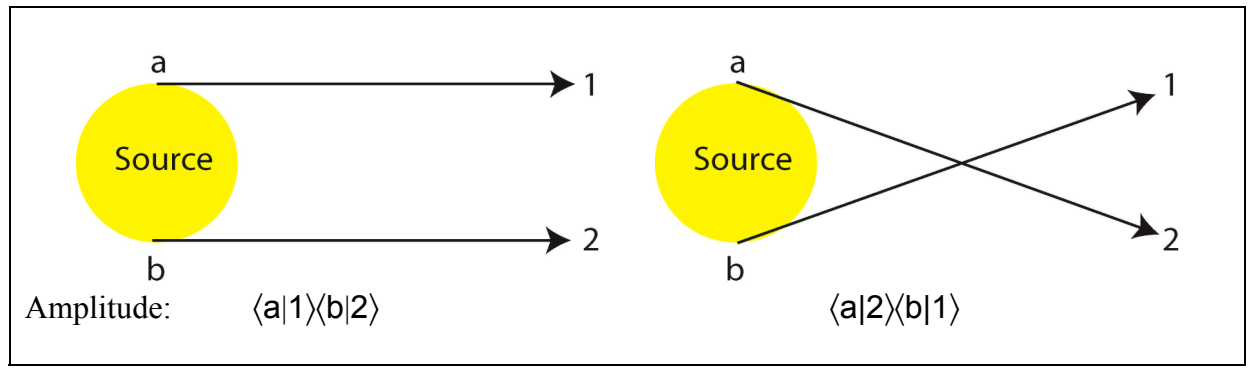

Box 2. Quantum interpretation of intensity correlations. Two photons from two source points, *a* and *b* are detected at points 1 and 2 following the two possible pairs of paths shown above. Each pair of paths corresponds to a quantum amplitude, and these amplitudes are added to find the probability to detect 2 photons:

$$P = |\langle a|1\rangle\langle b|2\rangle + \langle a|2\rangle\langle b|1\rangle|^2.$$

The interference term between these two amplitudes is washed out by the average over all the points in the source unless the distance between 1 and 2 is sufficiently small. This distance is the diffraction spot of the source. It is important to note that at the quantum level, the interference takes place in an abstract space in which the two photons are described as a single system. This explains the conceptual difficulty of the quantum interpretation. It is an example in which a simple classical effect obscures a rather subtle quantum effect. In a laser all the photons are in the same field mode: the points a and b are thus one and the same and there is no interference term.

In the case of fermions, the antisymetrization under exchange of the particles leads to a minus sign between the two amplitudes. Thus the probability of detecting two fermions at the same point at the same time is zero. Such anti-correlated fluctuations have no explanation in terms of speckle nor any other classical interpretation, be it wave-like or particle-like.

## The atomic Hanbury Brown - Twiss effect

The possibility of demonstrating an analogous effect with atoms has fascinated researchers at least since the emergence of atom optics in the 1990's, if not much earlier. Although one can easily generalize Glauber's formalism to atoms, one must remember that the HBT effect is particularly surprising when one interprets it in terms of particles, something which is not really necessary for light. Atoms on the other hand really seem like particles to us, and if they exhibit bunching, one can still, like Einstein, be troubled by the "mutual influence" between particles which arises without any force.

In addition atom optics permits one to ask new questions. What happens with fermions? The theoretical answer is unambiguous: the two amplitudes which interfere constructively for bosons must interfere destructively for fermions, and joint detection of indistinguishable fermions is less probable than that of independent particles – this effect is called anti-bunching and can be viewed as a manifestation of the exclusion principle. An interesting point is that no classical interpretation exists for anti-bunching, and as such one can say that fermions are "more" quantum mechanical than bosons, many of whose properties do have classical interpretations.

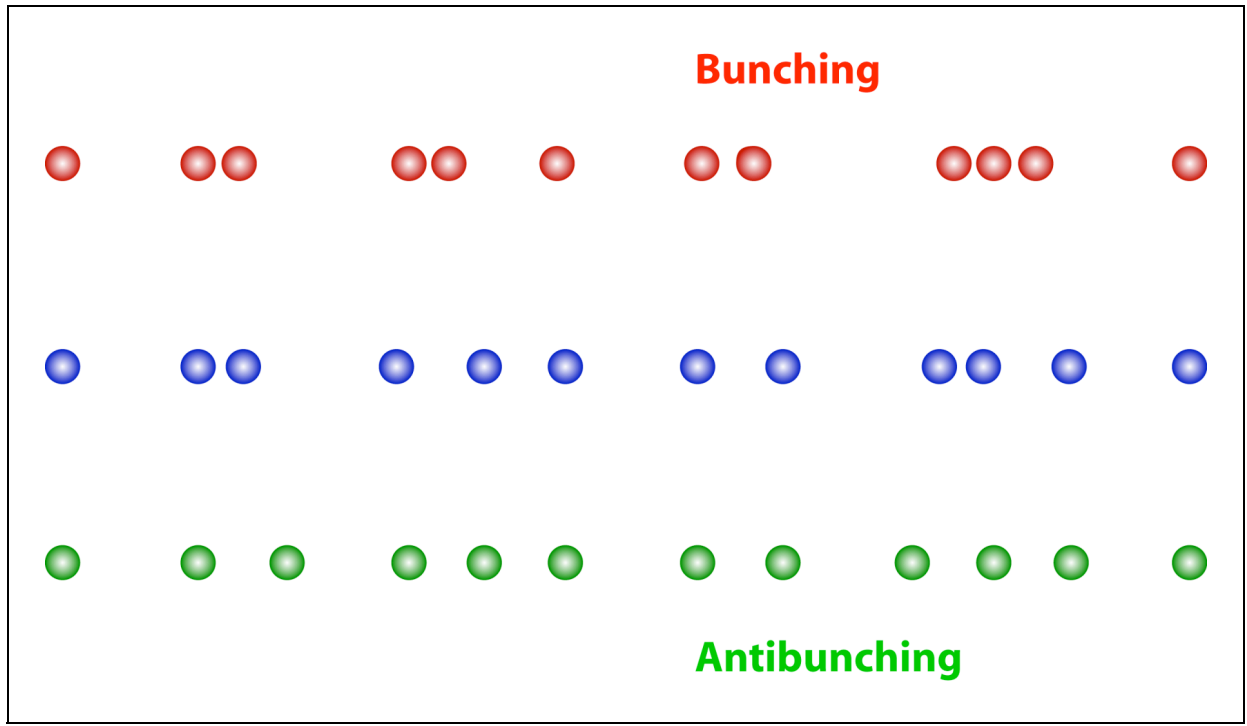

Box 3. The figure above illustrates the meaning of the terms bunching and antibunching. Suppose the spheres in each row represent the positions of a group of particles along a line. (They could also represent the arrival times at a detector.)
- The middle row of spheres (blue) corresponds to normal, everyday, *independent* particles. The spheres were placed at random positions on a line and the position of each sphere is independent of all the others. Sometimes they are close together sometimes they are far apart.
- The upper row (red) corresponds to particles exhibiting *bunching*. Note that compared to the independent particles, the particles tend to clump together. They are more often found together than independent particles.
- The lower row (green) represents *antibunched* particles. In this case they are more evenly spaced than independent particles. Indeed, they are never found close together.

Bunching or antibunching can have any of several physical origins. An obvious cause is interactions between the particles: if the particles attract each other they will clump together (bunching), if they repel each other they will spread out (antibunching) It is important to realize however, that in the experiments we have carried out on helium atoms, such interactions are completely negligible and that the origin of (anti-) bunching is a rather more mysterious interference effect.

In practice, observing bunching or anti-bunching with atoms is difficult, in part because the probability of finding two particles in the same elementary phase space cell is very small. Only the advances in cooling atoms with lasers and by evaporation over the past two decades have rendered the enterprise feasible. Lowering the temperature reduces the occupied volume in momentum space $(\Delta p)^3$. In addition, if the confining potential is curved (e.g. harmonic) rather than a box, lower temperature also corresponds to a smaller volume in real space $(\Delta x)^3$. Indeed, as one approaches the threshold for Bose-Einstein condensation, the average number of particles per phase space cell approaches unity and the signal to noise can be high. Unfortunately the real space volume over which the correlation is strong is quite small despite the low temperature, and it is difficult to achieve a resolution corresponding to a single phase space cell. The factor *g* is therefore greater that unity, decreasing the contribution of the second term in Eq. 1. In our experiment described below, *g* is of order 15. The contrast of the signal is diminished by a factor of this order.

**Experiment with metastable helium**

With these ideas in mind, our group at the Institut d'Optique has developed a detector capable of measuring spatio-temporal correlations between atoms by taking advantage of the properties of metastable helium. The metastable state $2^3S_1$ is 20 eV above the ground state and although its lifetime is 9000 s in vacuum, it rapidly deexcites when in contact with a metal surface (such as that of a microchannel plate) thus liberating 20 eV in the form of a free electron. The microchannel plate can thus provide an electrical pulse for a single atom. A position sensitive anode makes the plate the equivalent of an array of about $10^4$ separate detectors, capable of recording the arrival times and positions of a large number of atoms. The experimental set up is shown in Fig. 2. A cloud of cold, but not Bose condensed, atoms falls, accelerated by gravity, onto the detector. With the three dimensional arrival information of each atom, one can construct the correlation function by histogramming the number of detected atom pairs as a function of their separation. Typical data are shown in Fig. 3. Since all atoms arrive at the detector with essentially the same velocity, it is convenient to convert arrival times to vertical positions.

In 2005 we used helium-4 (a boson) to observe the atomic HBT effect with good signal to noise. In addition, by cooling the sample to below the Bose-Einstein condensation threshold, we were able to observe the absence of correlations in a BEC, illustrating the profound analogy between this state of matter and the light from a laser. More recently, in collaboration with a group in Amsterdam, we have also made the same measurement for helium-3, a fermion[6]. Indeed in that experiment we were able to study helium-3 and helium-4 clouds under nearly identical conditions in the same apparatus. The clouds were of sufficiently low density that the comparison of the two isotopes in Fig. 3 shows a purely quantum statistical effect. For thermal bosons the probability to detect two particles is increased while for fermions it is decreased. The spatial scale of the correlation can be understood with a calculation similar to that in the case of light (see Box 1); after a time of flight *t*, the correlation length is $ht/2\pi ms$, where *m* is the atomic mass, *h* is Planck's constant and *s* is the source size. The correlation length for light, $\lambda L/2\pi s$ can be recovered by identifying $h/mv$ with the de Broglie wavelength of an atom moving at speed $v = L/t$. If the detector has arbitrarily good resolution one expects a value 0 for fermions and 2 for bosons at zero separation. In the experiment the amplitude of the signal is limited by the detector resolution resulting in a factor *g* is of order 15 as mentioned above.

We have given a discription of our experiments at the Institut d'Optique, but many other groups have performed related experiments in recent years. In the references below we cite some of these experiments[7].

**Prospects**

In a sense, the correlations between atoms that we have observed are a consequence of elementary quantum theory, and one might ask why we and other researchers have gone to such efforts to observe them. One answer is that the experimental demonstration of non-trivial quantum phenomena, even elementary ones, often stimulates fruitful new ideas. But we also know that the demonstration of these correlations is only the beginning. The interaction between atoms, which we entirely neglect here, the possible formation of molecular dimers or the use of more complex configurations – such as putting the atoms in optical lattices – should lead to a rich variety of phenomena. Experiments similar to ours have been done on both bosons and fermions in optical lattices[8]. We know that in such structures, confined in one or two dimensions or in rotating traps, ultra cold atoms constitute important testing grounds for models from condensed matter physics. Some even hope to shed light on phenomena which are still poorly understood such as high temperature superconductivity.

Hanbury Brown - Twiss experiments have also been performed on other types of particles. In nuclear physics, correlations between pions give information about collision volumes in heavy ion collisions, thus realizing at the scale of femtometers a measurement analogous to HBT's measurements of stellar diameters[9]. In a solid, conduction electrons form an essentially ideal fermi gas and HBT type anti-correlations have been observed[10]. As in the case of atoms, these experiments are opening interesting paths towards the study of electron correlations in more exotic situations such as the fractional quantum Hall effect.

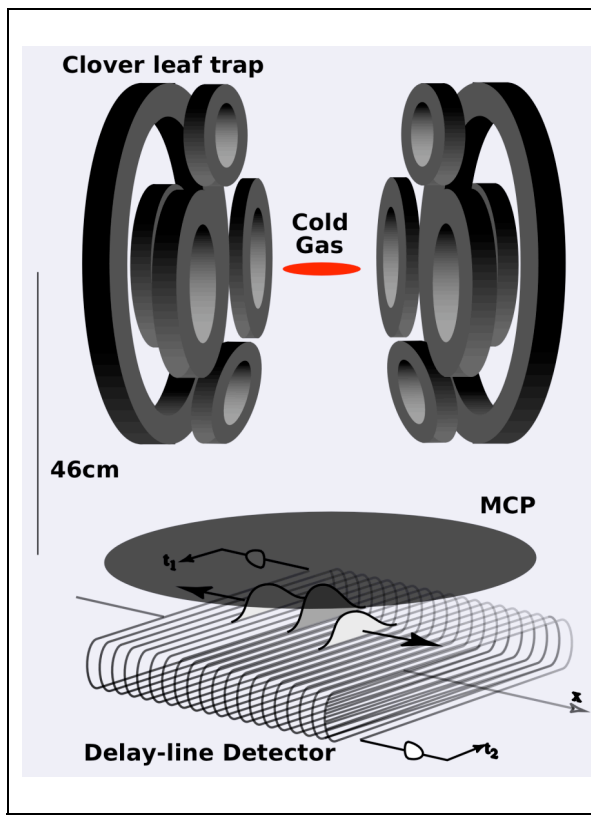 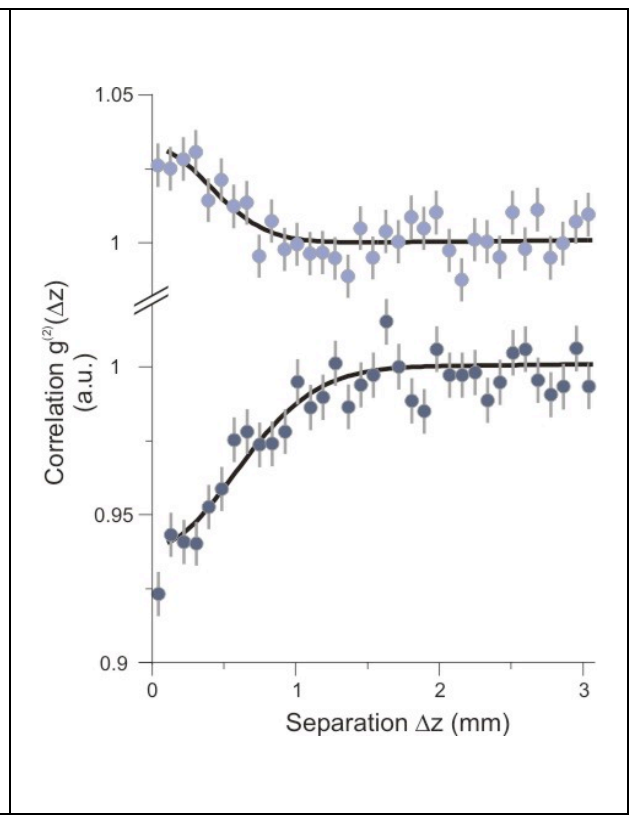

Figure 2. Experimental setup at the Institut d'Optique. The atoms are held in a magnetic trap where laser and evaporative cooling produce a cloud of $10^5$ atoms at a temperature of order one microKelvin. The detector is 80 mm in diameter and uses a system of delay line anodes to achieve a spatial resolution of about 0.5 mm.

Figure 3. Correlation function for helium-4 (bosons, upper curve) and helium-3 (fermions, lower curve). The bosons exhibit bunching, the fermions anti-bunching. Since the detector resolution is somewhat larger than the correlation length, the signal is spread out and the amplitude is reduced by a factor $g \approx 15$.